\newcommand{\ur}{{\bf r}}
\newcommand{\uR}{{\bf R}}

\newcommand{\uD}{{\bf D}}
\newcommand{\ux}{{\bf x}}
\newcommand{\uF}{{\bf F}}
\newcommand{\uk}{{\bf k}}

\newcommand{\uT}{{\bf T}}

\newcommand{\ua}{{\bf a}}
\newcommand{\uv}{{\bf v}}

\renewcommand{\Re}{{\sf Re}}
\renewcommand{\Im}{{\sf Im}}
\newcommand{\Heaviside}{\theta}

\documentstyle[12pt]{article}

\begin{document}

\title{{\bf Methods for molecular dynamics with nonadiabatic transitions}}
     
\author{D.F. Coker$^\dagger$ and L. Xiao,\\
\\
Department of Chemistry, Boston University,\\
590 Commonwealth Avenue, Boston, MA 02215\\}

\date{\ }

\maketitle

\begin{abstract}
We show how the dynamically nonlocal formulation of classical nuclear motion
in the presence of quantal electronic transitions presented many years ago by
Pechukas \cite{Pechukas1,Pechukas2} can be localized in time using time
dependent perturbation theory to give an impulsive force which acts when
trajectories hop between electronic surfaces. The action of this impulsive
force is completely equivalent to adjusting the nuclear velocities in the
direction of the nonadiabatic coupling vector so as to conserve energy, a
procedure which is widely used in surface hopping trajectory methods
\cite{Tully3}. This is the first time the precise connection between these
two formulations of the nonadiabatic dynamics problem has been considered.

We also demonstrate that the stationary phase approximation to the reduced
propagator at the heart of Pechukas' theory is not unitary due to its neglect
of nonstationary paths.  As such mixed quantum-classical evolution schemes
based on this approximation are not norm conserving and in general must fail
to give the correct branching between different competing electronic states.
Tully's phase coherent, fewest switches branching algorithm is guaranteed to
conserve the norm. The branching between different alternatives predicted by
this approach, however, may be inaccurate, due to use of the approximate local
dynamics. We explore the relative merits of these different approximations
using Tully's 1D two state example scattering problems for which numerically
exact results are easily obtained.
\end{abstract}

$^\dagger$ Presidential Young Investigator

\newpage

\section{Introduction}
\label{sec:intro}
\setcounter{equation}{0}
\baselineskip=24pt
\parindent=0.25in

There has recently been considerable interest in the development of
nonadiabatic molecular dynamics methods for describing condensed phase 
systems
which incorporate the possibility of transitions between electronic states
driven by classical nuclear motions
\cite{Tully3,Space1,Space2,Coker12,Webster1,Webster2,Webster4}
\cite{Kuntz1,Kuntz2,Gersonde1,Sawada1}.
One of the goals of this work is to develop microscopic techniques for
exploring the influence of solvent motions on chemical reactions which involve
changes in electronic state of the reactants.

The aim of these developments is to devise an accurate way to combine
classical dynamics methods to describe most of the degrees of freedom in a
problem, with quantum techniques to describe the remaining small number of the
degrees of freedom which we cannot avoid treating with some sort of quantal
representation. There is a long rich history of formal developments in this
area going back to early treatments of scattering
\cite{Miller5,Pechukas1,Pechukas2,Nikitin3,Child1,Penner1,Penner2,Billing1}. 
The advent of surface hopping methods \cite{Tully1,Tully2,Miller2} paved the
way for applications of these ideas to complicated realistic scattering
systems \cite{Stine1,Stine2,Stine3,Blais1,Blais2} but such methods are only
fairly accurate at high collision energies and most condensed phase
applications involve considerably smaller energies where these methods are
often unreliable.

Pechukas \cite{Pechukas1,Pechukas2} showed that the formal coupling of quantal
and classical subsystems resulted in equations of motion for the classical
degrees of freedom with forces that were nonlocal in time. Most existing
surface hopping techniques ignore this nonlocality and only recently
\cite{Tully3,Webster4} have these algorithms evolved to an extent where
they take proper account of phase coherent evolution of the quantal subsystem.

In this paper we explore the relationship between phase coherent local surface
hopping trajectories and the paths through phase space which are the solutions
of Pechukas' nonlocal equations of motion for some simple one dimensional test
problems.  Section \ref{subsec:semic} first summarizes the theory of
semiclassical evolution of wavefunctions in terms of Pechukas' stationary
phase approximation to the reduced propagator describing evolution of nuclear
wavefunctions over coupled electronic surfaces. In section
\ref{subsec:mixed_q-c} we find stationary phase approximate expressions for
the terminal occupation probabilities of different electronic states given an
initial nuclear wavefunction localized on some starting electronic state. This
stationary phase development motivates the mixed quantum-classical
representation which provides the formal basis for surface hopping methods.
This section also notes the norm conservation problem of Pechukas' stationary
phase path formulation. When the reduced propagator is approximated as a sum
over contributions from only stationary phase paths rather than including
contributions from all possible paths, the propagator is no longer unitary so
the semiclassical electronic occupation probabilities predicted by this theory
need not sum to unity. Next, in section \ref{subsec:tully} we summarize
Tully's phase coherent surface hopping dynamics algorithm and present a formal
derivation of the velocity rescaling force which acts on surface hopping
trajectories when systems switch between adiabatic electronic basis states.
The impulsive velocity rescaling force is shown to be the Pechukas force
localized in time using time dependent perturbation theory.

The accuracy of these different mixed quantum-classical dynamics methods is
studied by comparison of results with full quantum wave packet propagation
calculations for some simple one dimensional two state scattering problems in
section \ref{sec:results}. We also explore the norm conservation problems of
Pechukas' formulation and study the effects of temporal localization of the
force in surface hopping methods in these test problems.  Finally in section
\ref{sec:conclusion} we conclude with some ideas for reformulating these 
methods to address the problems which we have identified.

\section{Theory}
\label{sec:theory}
\setcounter{equation}{0}
\baselineskip=24pt
\parindent=0.25in

\subsection{Semiclassical Evolution of Wavefunctions}
\label{subsec:semic}

We consider a system with nuclear degrees of freedom $\uR$ and electronic
degrees of freedom $\ux$. The dynamics of the full system quantum wave
function $\psi(\ux,\uR,t)$ can be expressed in terms of the system propagator
$K(\ux'',\uR'',t''|\ux',\uR',t')$ as
\begin{equation}
\psi(\ux'',\uR'',t'') = \int d\ux' d\uR' K(\ux'',\uR'',t''|\ux',\uR',t') 
\psi(\ux',\uR',t')
\end{equation}
or writing the system wave function in terms of some basis set of electronic
states $\{\phi_n(\ux)\}$ thus 
\begin{equation}
\psi(\ux',\uR',t') = \sum_n \chi_n(\uR',t') \phi_n(\ux',t')
\end{equation}
the dynamics of the nuclear wave functions $\chi_n(\uR',t')$ on the different
electronic states is determined by the reduced propagator $K_{\beta \alpha}$
\cite{Pechukas1,Xiao1} as
\begin{equation}
\chi_{\beta}(\uR'',t'') = \sum_{\alpha} \int d\uR' K_{\beta
\alpha}(\uR'',t''|\uR',t') \chi_{\alpha}(\uR',t') \label{eq:reduce_evolve}
\end{equation}
$K_{\beta \alpha}(\uR'',t''|\uR',t')$ gives the probability amplitude for the
quantal system to change from state $\alpha$ to state $\beta$ as the nuclei
move from $\uR'(t')$ to $\uR''(t'')$. The probability of being in electronic
basis state $\beta$ at time $t''$ is thus
\begin{equation}
P_{\beta}(t'') = \int d\uR'' |\chi_{\beta}(\uR'',t'')|^2 \label{eq:prob}
\end{equation}
and $\sum_{\beta} P_{\beta}(t'') = 1$

Pechukas \cite{Pechukas1,Pechukas2} shows that the reduced propagators in
Eq.(\ref{eq:reduce_evolve}) can be written as an integration over all paths
$\uR(t)$ connecting the specified endpoints, thus
\begin{equation} 
K_{\beta\alpha}(\uR'',t''|\uR',t') = \int_{\uR'(t')}^{\uR''(t'')} 
{\cal D}[\uR(t)] 
\exp[{i \over \hbar} S_{o}[\uR(t)]] T_{\beta\alpha}[\uR(t)] \label{eq:fullpath}
\end{equation}
Here $S_{o}[\uR(t)] = \int_{t'}^{t''} \frac{1}{2} M \dot{R}^2(t) dt$ and 
\begin{eqnarray}  
T_{\beta\alpha}[\uR(t)] & = & \langle \phi_{\beta}(t'')
|\exp[-{i \over \hbar}
\int_{t'}^{t''} {\cal H}_{el}(\uR(t)) dt]|\phi_{\alpha}(t')\rangle \nonumber \\
& = & \langle \phi_{\beta}(t'')| \psi^{\alpha}(t'',t')\rangle 
\label{eq:tba}
\end{eqnarray}
is the transition amplitude of the quantal subsystem obtained by propagating
the initial instantaneous basis state $|\phi_{\alpha}(t') \rangle$ using 
the time dependent quantal Hamiltonian ${\cal H}_{el}(\ux,\uR(t))$ computed 
along the path $\uR(t)$ to a mixed state $|\psi^{\alpha}(t'',t') \rangle$ and 
overlapping with the final instantaneous basis state $\langle
\phi_{\beta}(t'')|$.

Pechukas derives a semiclassical approximation for the reduced propagator by
assuming that the paths of the nuclear coordinates which make the most
important contributions to the path integral expression in
Eq.(\ref{eq:fullpath}) are the paths of stationary phase $\bar{\uR}(t)$ which
satisfy 
\begin{eqnarray}
M\ddot{\bar{\uR}}(t) & = & \uF_{\beta\alpha}^P(t) \nonumber \\
\ & = & -\Re\left\{
\frac{\langle\bar{\psi}^{\beta}(t,t'')|\nabla_{\bar{\uR}(t)}{\cal H}_{el}|
\bar{\psi}^{\alpha}(t,t')\rangle}
{\langle\bar{\psi}^{\beta}(t,t'')|\bar{\psi}^{\alpha}(t,t') 
\rangle }\right\}
\label{eq:pechmot}
\end{eqnarray}
The effective potential corresponding to the force in
Eq.(\ref{eq:pechmot}) is 
\begin{equation}
{\cal V}_{eff}(t) = \Re\left\{
\frac{\langle\bar{\psi}^{\beta}(t,t'')|{\cal H}_{el}(t)|
\bar{\psi}^{\alpha}(t,t')\rangle}
{\langle\bar{\psi}^{\beta}(t,t'')|\bar{\psi}^{\alpha}(t,t') 
\rangle }\right\}
\label{eq:veff}
\end{equation}
Here the bars above the wavefunctions indicate that they are the solution of
the time-dependent Schr\"{o}dinger equation with hamiltonian ${\cal
H}_{el}(\ux,\bar{\uR}(t))$. Thus $\bar{\psi}^{\alpha}(t,t')$ is the mixed
state wavefunction at time $t$ which started out at time $t'$ in basis state
$\phi_{\alpha}(t')$, and $\bar{\psi}^{\beta}(t,t'')$ started out at time $t''$
in basis state $\phi_{\beta}(t'')$ and the time dependent quantal subsystem
hamiltonian evaluated along the stationary phase path ${\cal H}_{el}
(\ux,\bar{\uR}(t))$ has been used to propagate these boundary states forward
and backward in time to $t$. Pechukas shows that these nonadiabatic equations
of motion conserve both energy (using the effective potential) and angular
momentum.

By considering second order variations in the phase of $K_{\beta\alpha}$
Pechukas obtains the following semiclassical approximation for the reduced
propagator
\begin{eqnarray}
K^{SC}_{\beta \alpha}(\uR'',t''|\uR',t') & = & \sum_p\left({M \over 2 \pi i
\hbar}\right)^{3N/2} \left|\left({\partial \uR'' \over \partial
\dot{\uR}'}\right)_p\right|^{-1/2} \nonumber \\
& \ & \exp[i(S_o[\uR_p(t)]/\hbar - n_p \pi /2)] T_{\beta \alpha}[\uR_p(t)] 
\label{eq:scprop}
\end{eqnarray}
Where we have dropped the bars over the Pechukas stationary phase paths for
the sake of notational simplicity. With in the semiclassical approximation in
Eq.(\ref{eq:scprop}) the sum over only ``classical'' or stationary phase paths
of the nuclear coordinates, $\uR_p(t)$, from $\uR'(t')$ to $\uR''(t'')$ which
are solutions of Eq.(\ref{eq:pechmot}) replaces the integration over all
possible paths in the exact expression in Eq.(\ref{eq:fullpath}).  Further in
Eq.(\ref{eq:scprop}), $\left|\left(\partial \uR'' / \partial
\dot{\uR}'\right)_p\right|$ is the Jacobian of final positions with respect to
initial velocities of the stationary phase path $\uR_p(t)$, and $n_p$ is the
number of times the determinant $\left|\left(\partial \uR(t) / \partial
\dot{\uR}'\right)_p\right|$ is zero on the interval $t'<t<t''$ along the path
$\uR_p(t)$.

The Pechukas equations of motion, Eq.(\ref{eq:pechmot}), must be solved
iteratively due to the temporal nonlocality of $\uF_{\beta \alpha}^P(t)$ {\em
i.e.} the force on the trajectory at time $t$ depends on the forward and
backward propagated wavefunctions which can only be determined if we know the
full trajectory $\bar{\uR}(t)$. Pechukas suggests that we start with a guessed
trajectory thus specifying ${\cal H}_{el}(t)$, integrate the time-dependent
Schr\"{o}dinger equation forward and backward from the appropriate boundary
states and use the mixed state solutions, $\bar{\psi}^{\alpha}(t,t')$, and
$\bar{\psi}^{\beta}(t,t'')$ to determine the force in Eq.(\ref{eq:pechmot}),
which we solve for a new trajectory. This iterative approach provides the
basis of the algorithm recently reported by Webster and co-workers
\cite{Webster1,Webster2,Webster4}.

We interpret the above results as follows: If electronic state $\alpha$ is
initially occupied, the initial system wavefunction vector has components
$\chi_i(\uR',t) = 0$ $\forall \: i \ne \alpha$ and $\chi_{\alpha}(\uR',t) =
f(\uR')$, where $f(\uR')$ is the shape of the initial nuclear wavefunction on
electronic surface $\alpha$.  Substituting this initial wavefunction vector
into Eq.(\ref{eq:reduce_evolve}) the wavefunction vector at time $t''$ will
have components
\begin{equation}
\chi_{\beta}(\uR'',t'') = \int d\uR' K_{\beta \alpha}(\uR'',t''|\uR',t') 
f(\uR')
\label{eq:newchi} 
\end{equation}
Using $K_{\beta \alpha}^{SC}$ in Eq.(\ref{eq:newchi}) in place of $K_{\beta
\alpha}$ gives that the semiclassical approximation for the advanced nuclear 
wavefunction at $\uR''$ moving over electronic state $\beta$ has a
contribution from each initial nuclear position $\uR'$ which is just $f(\uR')$
multiplied by a sum of factors in Eq.(\ref{eq:scprop}) computed along each of
the Pechukas stationary phase paths which connect the initial position $\uR'$
to the final point $\uR''$ where we wish to evaluate the new wavefunction.
These stationary phase paths can in principle be obtained by means of a root
search due to the fact that they are specified in terms of a boundary value
problem.  Thus we scan the initial values of velocity $\dot{\uR}'$ and solve
Eq.(\ref{eq:pechmot}) iteratively till we find all converged paths which reach
the desired end point $\uR''$.  The final value of the new wavefunction is
obtained by integrating over these contributions from each initial point as in
Eq.(\ref{eq:newchi}).

Replacing $K_{\beta \alpha}$ with $K_{\beta \alpha}^{SC}$ as outlined above
provides a multidimensional definition of semiclassical mechanics akin to the
JWKB formulation. This semiclassical approach is somewhat impractical and the
mixed quantum-classical methods which we shall now describe provide a
viable alternative.

\subsection{State Occupation Probabilities and Mixed Quantum-Classical
Dynamics} 
\label{subsec:mixed_q-c}

Our interest is in experiments in which the system starts out in some initial
excited electronic state $\phi_{\alpha}(t')$, moves through a region where
there is strong dynamical mixing between quantum states, and emerges into a
mixture of final states, $\{\phi_{\beta}(t'')\}$, whose composition does not
change in time. This picture describes even condensed phase excited state
relaxation processes for which the interaction region may be complicated and
passage through it may take a considerable time.

We consider using Eq.(\ref{eq:prob}) and the above semiclassical results to
evaluate $P^{SC}_{\beta \alpha}(t'')$, the probability of terminating in state
$\phi_{\beta}(t'')$ after starting out in state $\phi_{\alpha}(t')$ with the 
example initial nuclear wavefunction
\begin{equation}
f(\uR') = \left(\pi \sigma^2 \right)^{-3N/4} \exp[-|\uR' - \uR_0|^2/2\sigma^2]
\exp[i \uk \cdot (\uR' - \uR_{0})] \label{eq:psi_init}
\end{equation}
Thus using Eqs.(\ref{eq:scprop}) - (\ref{eq:psi_init}) in Eq.(\ref{eq:prob})
we find 
\begin{eqnarray}
P^{SC}_{\beta \alpha}(t'') & =  & \left({M \over 2 \pi \hbar} \right)^{3N} 
\left(\pi \sigma^2 \right)^{-3N/2}\int d\uR''\int d\uR'\int d\uR'''  
\sum_b \sum_f \left|\left({\partial \uR'' \over \partial
\dot{\uR}'''}\right)_b \right|^{-1/2} 
\left|\left({\partial \uR'' \over \partial
\dot{\uR}'}\right)_f\right|^{-1/2} \nonumber \\
&  & \exp[i(S_o[\uR_f(t)]-S_o[\uR_b(t)])/\hbar] \exp[-i \pi (n_f-n_b)/2]
\label{eq:prob_sc} \\
& \  & T_{\beta \alpha}[\uR_f(t)] T^*_{\beta \alpha}[\uR_b(t)] 
\exp[i \uk \cdot (\uR' - \uR''')] 
\exp[-(|\uR''' - \uR_0|^2 + |\uR' - \uR_0|^2)/2\sigma^2] \nonumber
\end{eqnarray}
A diagramatic representation of this integral is presented in Fig.
\ref{fig:int_paths} where we see that the forward and backward paths arising 
from the absolute value in Eq.(\ref{eq:prob}) each start at different initial
points $\uR'$ and $\uR'''$ but both must terminate at $\uR''$, the point where
we evaluate the density, and over which we finally integrate.

The phase factor in the integrand of the above expression has the form
$\exp[i\Phi/\hbar]$ with 
\begin{eqnarray}
\Phi & = & (S_o[\uR_f(t)] - S_o[\uR_b(t)]) - \hbar \pi (n_f - n_b) / 2 + 
\hbar \uk \cdot (\uR' - \uR''') \nonumber \\
 &  & + \hbar \Im\{ \ln T_{\beta \alpha}[\uR_f(t)] \} 
+ \hbar \Im\{ \ln T_{\beta \alpha}^*[\uR_b(t)] \} \label{eq:phase}
\end{eqnarray}
This phase is a functional of the forward and backward paths $\uR_f(t)$ and
$\uR_b(t)$ respectively and as such it is a function of the independent
variables, the path endpoints $\uR'$, $\uR''$, and $\uR'''$. The most
important contributions to the integrals over endpoints will come when the
phase is stationary with respect to variations in these path endpoints. We
consider a generalized variation of the paths about a stationary phase point
as indicated in Fig. \ref{fig:int_paths} which involves adding variation paths
$\delta \uR_f(t)$ and $\delta \uR_b(t)$ to both the forward and backward
paths. The new paths are thus obtained by varying the initial points $\uR'$
and $\uR'''$ by amounts $\delta \uR_f(t') = \delta \uR'$ and $\delta \uR_b(t')
= \delta \uR'''$ respectively and varying the common terminal point $\uR''$ by
$\delta \uR'' = \delta \uR_f(t'') = \delta \uR_b(t'')$. Given the new
endpoints we scan the forward and backward path initial momenta and solve 
the
Pechukas equations of motion till we find Pechukas paths which connect the new
endpoints. The different components of the phase variation resulting from such
path displacements taking explicit account of endpoint variations are readily
evaluated as
\begin{equation}
\delta S_0[\uR(t)] = M \dot{\uR}(t'') \delta \uR(t'') - M \dot{\uR}(t') \delta
\uR(t') - \int_{t'}^{t''} M \ddot{\uR}(t) \delta \uR(t) dt 
\end{equation}
and 
\newpage
\begin{eqnarray}
\delta \hbar \Im \{ \ln T_{\beta \alpha}[\uR(t)] \} & = & 
\Im \{ {\hbar \over T_{\beta \alpha}} [ \langle \nabla_{\uR} 
\phi_{\beta}(t'') | \exp[{-i \over \hbar}\int_{t'}^{t''} 
{\cal H}_{el}(t) dt] | \phi_{\alpha}(t') \rangle \delta \uR(t'') \nonumber \\
\ & \ & + \langle \phi_{\beta}(t'') | \exp[{-i \over \hbar}\int_{t'}^{t''} 
{\cal H}_{el}(t) dt] | \nabla_{\uR} \phi_{\alpha}(t') \rangle \delta \uR(t')
] \} \\
\ & \ & - \Re \left \{ {1 \over T_{\beta \alpha}} \langle \phi(t'') |
\int_{t'}^{t''}dt \delta \uR(t) \nabla_{\uR} {\cal H}_{el}(t)  
\exp[{-i \over \hbar}\int_{t'}^{t''} 
{\cal H}_{el}(t) dt] | \phi_{\alpha}(t') \rangle \right \}
\nonumber
\end{eqnarray}
Using these results we find that the variation of the phase in
Eq.(\ref{eq:phase}) arising from the full generalized path variation discussed
above is
\begin{eqnarray}
\delta \Phi & = & \delta \uR' \left[ -M \dot{\uR}_f(t') + \hbar \left( \uk + 
\Im \left \{{\sum_n T_{\beta n}[\uR_f(t)] \langle \phi_n(t')| \nabla_{\uR}
\phi_{\alpha}(t') \rangle_{\uR_f} \over T_{\beta \alpha}[\uR_f(t)] } 
\right \} \right) \right]  \nonumber \\
\ & - & \delta \uR'''\left[ -M \dot{\uR}_b(t') + \hbar \left( \uk + 
\Im \left \{{\sum_n T_{\beta n}[\uR_b(t)] \langle \phi_n(t')| \nabla_{\uR}
\phi_{\alpha}(t') \rangle_{\uR_b} \over T_{\beta \alpha}[\uR_b(t)] } 
\right \} \right)\right]  \\
\ & + & \delta \uR'' [ M(\dot{\uR}_f(t'') - \dot{\uR}_b(t'')) \nonumber \\
\ & \ & + \hbar \Im \left \{
\sum_n \left ( {T_{n \alpha}[\uR_f(t)] \langle \nabla_{\uR} \phi_{\beta}(t'')| 
\phi_n(t'') \rangle_{\uR_f} \over T_{\beta \alpha}[\uR_f(t)] } \right )
- \left ( {T_{n \alpha}[\uR_b(t)] \langle \nabla_{\uR} \phi_{\beta}(t'')| 
\phi_n(t'') \rangle_{\uR_b} \over T_{\beta \alpha}[\uR_b(t)] } \right )
\right \} ]  \nonumber
\end{eqnarray} 
Here the terms with interior path variations vanish because we use solutions
to Pechukas' stationary phase equations of motion in our construction of
the variation, leaving only the above boundary terms \cite{Goldstein1}.

We require $\delta \Phi = 0$ for arbitrary independent endpoint variations,
$\delta \uR'$, $\delta \uR''$, and $\delta \uR'''$, thus phase stationarity
occurs with the following choice of boundary conditions
\begin{equation}
M \dot{\uR}_f(t') = \hbar \uk + 
\hbar \Im \left \{{\sum_n T_{\beta n}[\uR_f(t)] \uD_{n \alpha}(\uR_f(t'))
\over T_{\beta \alpha}[\uR_f(t)] } \right \} \label{eq:ic_forward}
\end{equation}
\begin{equation}
M \dot{\uR}_b(t') = \hbar \uk + 
\hbar \Im \left \{{\sum_n T_{\beta n}[\uR_b(t)] \uD_{n \alpha}(\uR_b(t'))
\over T_{\beta \alpha}[\uR_b(t)] } \right \} \label{eq:ic_back}
\end{equation}
and 
\begin{equation}
M(\dot{\uR}_f(t'') - \dot{\uR}_b(t'')) =
\hbar \Im \left \{
\sum_n \left ( {T_{n \alpha}[\uR_f(t)] \uD_{n \beta}^*(\uR_f(t''))
\over T_{\beta \alpha}[\uR_f(t)] } \right )
- \left ( {T_{n \alpha}[\uR_b(t)] \uD_{n \beta}^*(\uR_b(t''))
\over T_{\beta \alpha}[\uR_b(t)] } \right )
\right \} \label{eq:ic_connect}
\end{equation}
In the above results quantities like 
$\uD_{n \alpha}(\uR_f(t')) = \langle \phi_n(t')| 
\nabla_{\uR} \phi_{\alpha}(t') \rangle_{\uR_f}$ are the nonadiabatic coupling
vectors for the basis states at the appropriate endpoint of the specified
path. These quantities play a key role in the theory of nonadiabatic
processes. The stationary phase boundary conditions in the above equations are
in general nonlocal as the transition amplitudes, $T_{n \alpha}[\uR_f(t)]$
from initial state $\alpha$ to all possible final states $n$, for example, are
functionals of the entire specified path. If the path endpoints of interest
lie in regions where the nonadiabatic coupling vectors are small (asymptotic
scattering-like problems for example) we can ignore this initial nonlocality
and we thus find that phase stationarity
(Eq.(\ref{eq:ic_forward})-(\ref{eq:ic_back})) requires that the forward and
backward paths have the same initial momenta which is related to the
derivative of the phase of the initial wave function and for the wavefunction
considered above we thus have
\begin{equation}
M\dot{\uR}' \: \: = \: \: M\dot{\uR}''' \: \: = \: \: \hbar \uk
\end{equation}

For asymptotic endpoint conditions, Eq.(\ref{eq:ic_connect}) gives that the
final momenta of the forward and backward paths should also be equal to one
another {\em i.e.} $M\dot{\uR}_f(t'') = M\dot{\uR}_b(t'')$. In one dimension
the only way for two trajectories to start at points $\uR'$ and $\uR'''$ in
the asymptotic region with the same initial velocities and both terminate
after $T=t''-t'$ at the same point $\uR''$ in another asymptotic region, each
having the same terminal velocity is to have $\uR' = \uR'''$ and the forward
and backward paths are thus the same $\uR_f(t) = \uR_b(t)$. The situation is
obviously considerably more complex in many dimensions. In order to proceed,
however, we shall assume that the dominant contribution to the integrals over
endpoints consistent with these phase stationarity requirements comes from
paths for which $\uR' = \uR'''$. Using a reasonably well localized Gaussian
initial wavefunction will ofcourse damp out contributions from forward and
backward pairs of paths with the same endpoint velocities but with widely
separated initial positions thus aiding the validity of our assumption.

Incorporating the various requirements of phase stationarity discussed above
to limit the integrations and summations in Eq.(\ref{eq:prob_sc}) to their
most important contributions we thus have 
\begin{eqnarray}
P^{SC}_{\beta \alpha}(t'') & =  & \left({M \over 2 \pi \hbar} \right)^{3N} 
\left(\pi \sigma^2 \right)^{-3N/2}\int d\uR''\int d\uR'
\sum_p \left|\left({\partial \uR'' \over \partial
\dot{\uR}'}\right)_p \right|^{-1} |T_{\beta \alpha}[\uR_p(t)]|^2 \nonumber \\
&  & \exp[-|\uR' - \uR_0|^2/\sigma^2] \delta(\dot{\uR}' - {\hbar \uk / M}) 
\end{eqnarray}
Finally the inverse Jacobian and the sum over stationary phase paths can be
used to replace the integration over final positions $\uR''$ with an integral
over initial velocities \cite{Heller2,Heller3,Miller3,VanVleck1}. Thus the
stationary phase approximate expression for the occupation probability with
the Gaussian initial wavefunction in Eq.(\ref{eq:psi_init}) is finally
\begin{equation}
P^{SC}_{\beta \alpha}(t'') = \left(\pi \sigma^2 \right)^{-3N/2}\int d\uR' \int 
d\dot{\uR}'
\exp[-|\uR' - \uR_0|^2/\sigma^2] \delta(\dot{\uR}' - {\hbar \uk / M}) 
|T_{\beta \alpha}[\uR(t)]|^2
\label{eq:prob_fin} 
\end{equation}
where the stationary phase paths $\uR(t)$, along which we must evaluate the
transition amplitude $T_{\beta \alpha}$, are converged solutions of
Eq.(\ref{eq:pechmot}) which start at the different possible initial
points $\uR'$ but they all have the same initial momentum $M\dot{\uR}' = \hbar
\uk$.

%It is interesting to note that the mixed quantum-classical propagation
%proceedure implied by the stationary phase result in Eq.(\ref{eq:prob_fin})
%does not permit wavepacket spreading over a flat potential surface as the
%trajectories started at different points in space all have the same initial
%velocity thus the shape of the packet over a flat surface will not change as a
%function of time. Quantum-mechanically ofcourse a packet of the form in
%Eq.(\ref{eq:psi_init}) will spread over a flat potential as time goes on due
%to the inclusion of different momentum paths which make nonstationary
%contributions to the integral in Eq.(\ref{eq:prob_sc}). The mixed
%quantum-classical dynamics based on the stationary phase trajectories
%described above doesnot allow for such spreading. In the condensed phase
%collisions with surrounding molecules generally tend to localize nuclear
%wavefunctions and prevent spreading so this approximation should not
%present a
%problem for such applications.

Expectation values of nuclear operators for different electronic states are
readily obtained from this mixed quantum-classical formulation using the
result
\begin{equation}
\langle \hat{O} \rangle^{\alpha}_{\beta}(t'')= {\int d\uR'' 
\chi^*_{\beta}(t'') 
\hat{O} \chi_{\beta}(t'') \over \int d\uR'' \chi^*_{\beta}(t'') 
\chi_{\beta}(t'')}
\label{eq:gen_av}
\end{equation}
which readily gives the mean nuclear position, for example, as
\begin{equation}
\langle \uR\rangle^{\alpha}_{\beta}(t'')= { \int d\uR' \int d\dot{\uR}'
\exp[-|\uR' - \uR_0|^2/\sigma^2] \delta(\dot{\uR}' - {\hbar \uk / M}) 
|T_{\beta \alpha}[\uR(t)]|^2 \uR''(\uR',\dot{\uR}')
\over \int d\uR' \int d\dot{\uR}'
\exp[-|\uR' - \uR_0|^2/\sigma^2] \delta(\dot{\uR}' - {\hbar \uk / M}) 
|T_{\beta \alpha}[\uR(t)]|^2} \label{eq:R_av}
\end{equation}

The mixed quantum-classical representation, motivated by the stationary phase
developments discussed above involves describing the nuclear degrees of
freedom by a point in phase space $(\uR,\dot{\uR})$, and the electronic
degrees of freedom are represented by the amplitudes $a_n^{\alpha}(t,t')$ or
$b_n^{\beta}(t,t'')$ of the different instantaneous basis states. Here the
forward and backward propagated mixed state wavefunctions are
\begin{equation}
|\psi^{\alpha}(t,t')\rangle = \sum_n a_n^{\alpha}(t,t') |\phi_n(t) \rangle
\end{equation}
and 
\begin{equation}
\langle \psi^{\beta}(t,t'')| = \sum_m b_m^{\beta}^*(t,t'') 
\langle \phi_m(t) | \label{eq:ad_expand}
\end{equation}
We thus replace the notion of a nuclear wave function moving over electronic
surfaces with a swarm phase space trajectories initiated from different points
according to some initial phase space density distribution as in
Eq.(\ref{eq:prob_fin}). According to this equation, the square of the final
amplitude $|T_{\beta \alpha}[\uR(t)]|^2=|a_{\beta}^{\alpha}(t'',t')|^2$
computed for a particular trajectory gives the probability of the final
electronic state $\beta$ being occupied by this trajectory. Specifying a swarm
member's initial classical variables $(\uR_0,\dot{\uR}_0)$ and initially
occupied quantum basis state $\alpha$, together with Eq.(\ref{eq:pechmot})
uniquely determines a set of final phase space points
$\{(\uR_N,\dot{\uR}_N)_{\beta}\}$, one for each possible final electronic
state $\beta$.

Application of the mixed quantum-classical dynamics approach described above
is in principle straight forward: For each swarm member we first guess a phase
space trajectory represented as a discrete set of points displaced uniformly
in time starting at $(\uR_0,\dot{\uR}_0,t_0)$ and terminating at
$(\uR_N,\dot{\uR}_N,t_N)$ where $t_k = t'+k\Delta$ with $\Delta = T/N$ and $T
= t'' - t'$ is long on the timescale of nonadiabatic relaxation phenomenon of
interest.  The electronic Hamiltonian ${\cal H}_{el}(\ux,\uR_k)$ is evaluated
at each point along the path and a basis set to represent the electronic
states at each configuration is established. The instantaneous adiabatic basis
is often convenient.  Next we use ${\cal H}_{el}(\ux,\uR_k)$ to propagate the
initial state $\phi_{\alpha}(t')$ forward, and all possible final states
$\{\phi_{\beta}(t'')\}$ backward in time to give time histories of the mixed
states $\psi^{\alpha}$ and $\{\psi^{\beta}\}$ which are used to determine the
Pechukas forces in Eq.(\ref{eq:pechmot}). This equation is then integrated
away from $(\uR_0,\dot{\uR}_0,t_0)$ with each one of the different terminal
state Pechukas forces to give a set of new trajectories and the iteration
process outlined earlier is repeated till each phase space trajectory
terminating in each of the different relevant basis states $\{\phi_{\beta}\}$
is converged. This is done for all the swarm members under the initial phase
space density distribution.

For the asymptotic scattering-like problems we consider here any stationary
phase path $\bar{\uR}_{\beta \alpha}(t)$ starting in state $\alpha$ and
finishing in state $\beta$ with a given momentum will give the same value of
$T_{\beta \alpha}$ independent of initial and final positions in the
asymptotic regions.  Under these circumstances the stationary phase
approximation to the terminal state occupation probabilities are
$P^{SC}_{\beta \alpha} = |T_{\beta \alpha}|^2$.  The exact expression for 
$P_{\beta \alpha}$ is readily obtained
in the same way as the semiclassical result in Eq.(\ref{eq:prob_sc}) was
derived, only now using the full path integral expression for $K_{\beta
\alpha}$ in Eq.(\ref{eq:fullpath}) in place of the semiclassical approximation
of Eq.(\ref{eq:scprop}). The exact expression thus involves integrations over
all possible paths and the resulting values of $P_{\beta \alpha}$ will
ofcourse be different from those obtained when only stationary phase paths are
included. Nonstationary paths may interfere constructively or destructively so
the exact value of $P_{\beta \alpha}$ could be larger or smaller than the
semiclassical approximation. The full path integral expression for the reduced
propagator is unitary so the exact values of the occupation probabilities
conserve the norm and $\sum_{\beta} P_{\beta \alpha} = 1$ for all time. The
stationary phase approximation to the reduced propagator is not unitary and
since we neglect constructively or destructively interfering contributions
when we calculate $P^{SC}_{\beta \alpha}$, the norm is no longer guaranteed to
be conserved and it can be greater or less than unity.  We will explore this
serious problem in some detail in our test calculations presented in section
\ref{sec:results}.

\subsection{Surface Hopping Swarm Dynamics}
\label{subsec:tully}

The main practical problem with implementation of the mixed quantum-classical
dynamics method described in the previous subsection is the nonlocal nature of
the forces in the equations of motion for the stationary phase trajectories.
Surface hopping methods provide an approximate, intuitive, stochastic
alternative approach which uses the average dynamics of a swarm of
trajectories over the coupled surfaces to approximate the behavior of the
nonlocal stationary phase trajectories. With this approach the motion of each
swarm member is determined by completely local forces arising from a specially
selected instantaneous potential surface. As the swarm moves through a
coupling region its members partition between the competing surfaces
stochastically according to some Monte Carlo transition probability at each
step of the local dynamics.  There are thus two key aspects of a surface
hopping method: (1) We need a Monte Carlo rule for branching the members of
the trajectory swarm onto the different coupled surfaces so that the average
of the local swarm dynamics best approximates the nonlocal dynamics of the
Pechukas stationary phase trajectories, and (2) When a swarm member switches
from one surface to another in a nonadiabatic transition during a time step we
need a realistic model to describe how the energy redistributes between the
quantal and classical degrees of freedom.

Tully's phase coherent, fewest switches surface hopping algorithm
\cite{Tully3} is perhaps the most sophisticated of these methods. With this 
approach each ensemble member $i$, represented by its current nuclear position
and velocity $(\uR^{(i)}(t') \dot{\uR}^{(i)}(t'))$, and a set of expansion
coefficients $\{a^{(i)}_k(t')\}$, is always assumed to occupy only one of the
instantaneous basis states at a given time. This occupied basis state
determines the local force active on the trajectory at time $t'$. It is easily
shown by substituting Eq.(\ref{eq:ad_expand}) into the Schr\"{o}dinger
equation \cite{Nikitin3} that the coefficients satisfy 
\begin{equation}
{{\rm d} a_i \over {\rm d} t} = \sum_k d_{ik} a_k     \label{eq:amotion}
\end{equation}
where $d_{ik} = -i E_{ik}/\hbar - \langle \phi_i | \partial/\partial t |
\phi_k \rangle$ and $E_{ik} = \langle \phi_i | {\cal H}_{el} | \phi_k
\rangle$.
As we advance time through one step $\Delta$, the nuclei in ensemble member
$i$ move and its mixed electronic wavefunction changes composition. The Monte
Carlo sampling approach is employed to decide whether the ensemble member will
stay in its currently occupied state or if it will make a transition to some
other state during $\Delta$.  With Tully's approach each ensemble member makes
stochastic branches into the different instantaneous basis states
independently in such a way as to give a distribution of ensemble members
amongst the states consistent with the coherently propagated mixed electronic
state along the trajectories which are determined by completely local forces.

Tully's Monte Carlo sampling procedure is designed so that hops between
states only occur in regions where the expansion coefficients are changing
rapidly. Outside these regions, where the wave function is essentially a
stationary mixture of states, hopping ceases and each ensemble member stays in
its selected state. The algorithm uses the following Monte Carlo transition
probability to determine whether a transition from the occupied
state $\alpha$ to some other state $\beta$ should be made during the
interval $\Delta$
\begin{equation}
g_{\alpha \rightarrow \beta} =
{ 2 \Delta \Re\left(\rho_{\beta \alpha} d_{\beta \alpha}
\right) \over \rho_{\alpha \alpha} } 
\Heaviside\left( -\Re(\rho_{\beta \alpha}d_{\beta \alpha}) \right)
\label{eq:gitok}
\end{equation}
Here $\rho_{\beta \alpha}(t') = a_{\beta}^*(t') a_{\alpha}(t') = \rho_{\alpha
\beta}^*$, and $d_{\beta \alpha} = -d_{\alpha \beta}^*$ was defined in 
Eq.(\ref{eq:amotion}). The quantity $2\Re(\rho_{\beta \alpha}d_{\beta
\alpha})$ is the component of the flux of occupation probability out of state
$\alpha$ associated with transitions into state $\beta$ \cite{Tully3,Coker12}
and the Heaviside function $\Heaviside\left( -\Re(\rho_{\beta \alpha}d_{\beta
\alpha}) \right)$ permits only transitions which will depleat state $\alpha$'s
occupation if the sign of the flux is correct. This is Tully's ``fewest
switches'' Monte Carlo occupation sampling procedure and it can be derived by
setting up a master equation for the flux \cite{Tully3,Coker12}. It provides
rigorously norm conserving quantal subsystem dynamics for each swarm member
since we use expansion coefficients obtained by solving the time dependent
Schr\"{o}dinger equation along a single nuclear trajectory, rather than
different trajectories for the different possible transitions as in Pechukas'
formulation.

The final ingredient of Tully's algorithm is a prescription for determining
the effective potential over which the local dynamics of each ensemble member
takes place. This is done as indicated above by resolving the dynamical mixed
state evolving along each trajectory into a particular instantaneous basis
state to determine the local force on this trajectory. In this paper we adopt
a new approach by using time dependent perturbation theory \cite{Thirum10} 
to localize the nonlocal force in Pechukas' equation of motion,
Eq.(\ref{eq:pechmot}). The ideas here were motivated by the implementation of 
Pechukas' theory presented by Webster {\em et al.} \cite{Webster1}. Thus if we 
decide that some ensemble member changes
state from $\phi_{\alpha}(t') \rightarrow \phi_{\beta}(t'+\Delta)$ while the
nuclei move from $\uR(t') \rightarrow \uR(t'+\Delta)$ during the time step,
the Pechukas force operative on this trajectory at time $t$ lying in the range
$t' \le t \le t'+\Delta$ is obtained from Eq.(\ref{eq:pechmot}) as
\begin{equation}
\uF^P_{\beta \alpha}(t) = -\Re\left\{ { \sum_m \sum_n 
b_m^{\beta}^*(t,t'+\Delta) a_n^{\alpha}(t,t')
\langle \phi_m(t)|\nabla_{\uR} {\cal H}_{el}(t)| \phi_n(t) \rangle 
\over \sum_n b_n^{\beta}^*(t,t'+\Delta) a_n^{\alpha}(t,t') } \right\}
\label{eq:fp_adia}
\end{equation}
where the forward and backward propagated mixed state wavefunctions have 
been
written in terms of the instantaneous adiabatic basis states as in
Eqs.(\ref{eq:ad_expand}) and in this representation the effective potential
has the form
\begin{equation}
{\cal V}_{eff}(t) = \Re\left\{ {\sum_n 
b_n^{\beta}^*(t,t'+\Delta) a_n^{\alpha}(t,t') E_n(t)
\over \sum_n b_n^{\beta}^*(t,t'+\Delta) a_n^{\alpha}(t,t') } \right\}
\label{eq:veff_adia}
\end{equation}

In general we choose $\Delta$ relatively small so that appreciable changes in
the mixing coefficients can only result from the cumulative effect of several
steps. We employ this idea to avoid the use of the self consistent iteration
process and determine an effective Pechukas force which is local in time as
follows: Suppose that at the beginning of the time step the quantal subsystem
has been resolved into the instantaneous adiabatic state $\alpha$ thus
$a_n^{\alpha}(t',t') = 0$ for all $n \ne \alpha$ and
$a_{\alpha}^{\alpha}(t',t') = 1$. The Pechukas force active at the beginning
of the step from Eq.(\ref{eq:fp_adia}) is thus
\begin{equation}
\uF_{\beta \alpha}^P(t') = -\Re\left\{ {\sum_n b_n^{\beta*}(t',t'+\Delta) 
\langle \phi_n(t') | \nabla_{\uR} {\cal H}_{el}(t') | 
\phi_{\alpha}(t') \rangle
\over b_{\alpha}^{\beta *}(t',t'+\Delta) } \right\} \label{eq:fpexact}
\end{equation}
For the determination of the forces we assume that the quantum subsystem is
resolved into state $\beta$ at time $t'+\Delta$ thus the $b_n^{\beta}$
expansion coefficients could be obtained by back propagating the equations of
motion in Eq.(\ref{eq:amotion}) from the initial condition
$b_{\beta}^{\beta}(t'+\Delta,t'+\Delta) = 1$ and
$b_n^{\beta}(t'+\Delta,t'+\Delta) = 0$ for all $n \ne \beta$. If $\Delta$ is
small, back propagation will not change these coefficients too much from their
terminal values so $b_{\beta}^{\beta}(t',t'+\Delta) \sim 1$ and
$b_n^{\beta}(t',t'+\Delta) \sim 0$ for all $n \ne \beta$. With this time
dependent perturbation theory approximation the Pechukas force in
Eq.(\ref{eq:fpexact}) becomes strongly dominated by a single term and
\begin{eqnarray}
\uF_{\beta \alpha}^P(t') & \sim & -\Re\left\{ {b_{\beta}^{\beta*}
(t',t'+\Delta) \over b_{\alpha}^{\beta*}(t',t'+\Delta)} 
\langle \phi_{\beta}(t') | 
\nabla_{\uR} {\cal H}_{el}(t') | \phi_{\alpha}(t') \rangle \right\} 
\label{eq:fpapprox1} \\
 \ & = & \ -\Re\left\{ {b_{\beta}^{\beta*}(t',t'+\Delta) \over
b_{\alpha}^{\beta*}(t',t'+\Delta)} (E_{\beta} - E_{\alpha})\langle
\phi_{\beta}(t') | \nabla_{\uR} |
\phi_{\alpha}(t') \rangle \right\} \label{eq:fpapprox2} 
\end{eqnarray}
where we have used the fact that the force matrix elements are
\begin{equation}
\uF_{\beta \alpha} = -\langle \phi_{\beta}| \nabla_{\uR} {\cal H}_{el} | 
\phi_{\alpha} \rangle = 
(E_{\beta} - E_{\alpha}) \langle \phi_{\beta} | \nabla_{\uR}| \phi_{\alpha} 
\rangle \:\:\:\:\:\: \alpha \neq \beta 
\end{equation}
and
\begin{equation}
\uF_{\alpha \alpha} = -\langle \phi_{\alpha}| \nabla_{\uR} {\cal H}_{el} | 
\phi_{\alpha} \rangle = - \nabla_{\uR} E_{\alpha} \label{eq:fhell}
\end{equation}
which are easily obtained from the result $\nabla_{\uR} \langle \phi_{\beta} 
| {\cal H}_{el} | \phi_{\alpha} \rangle = \nabla_{\uR} E_{\alpha} 
\delta_{\beta \alpha}$. We note that if no transition occurs during $\Delta$
{\em i.e.} $\beta = \alpha$, Eq.(\ref{eq:fpapprox1}) together with
Eq.(\ref{eq:fhell}) give $\uF_{\alpha \alpha}^P(t') = -\nabla_{\uR}
E_{\alpha}$ as expected.

In a similar fashion, the equations of motion for the $b_n$ coefficients (we
drop the initial state $\beta$ superscript for simplicity) during the back
propagation
\begin{equation} 
{d b_k \over dt} = \sum_n d_{kn} b_n
\end{equation}
with initial conditions $b_{\beta}(t'+\Delta) = 1$, and $b_n(t'+\Delta) = 0$ 
for all $n \ne \beta$, decouple within this time dependent perturbation theory
approximation thus
\begin{eqnarray} 
{db_k \over dt} & = & d_{k \beta}b_{\beta} \:\:\:\:\:\: k \neq \beta 
\nonumber \\
{db_{\beta} \over dt} & = & d_{\beta \beta}b_{\beta} \label{eq:tdpert}
\end{eqnarray}
If we assume that the coupling matrix elements $d_{k \beta}=-i
E_{k \beta}/\hbar - \langle \phi_k| \partial / \partial t | \phi_{\beta} 
\rangle = -d_{\beta k}^*$ are independent of time over the interval 
$\Delta$ then Eqs.(\ref{eq:tdpert}) have solutions
\begin{eqnarray}
b_{\beta}(t) & = & \exp[d_{\beta \beta}(t-(t'+\Delta))] \nonumber \\
b_k(t) & = & \left( {d_{k \beta} \over d_{\beta \beta}}
\right)\left\{\exp[d_{\beta \beta}(t-(t'+\Delta))] - 1 \right\}
\end{eqnarray}
In the adiabatic representation we have that $d_{\beta \beta} = 
-i E_{\beta}/\hbar$ and $d_{k \beta} = - \langle \phi_k | \partial/\partial t|
\phi_{\beta} \rangle$ for all $k \ne \beta$, thus
\begin{equation}
{b_{\beta}^*(t') \over b_{\alpha}^*(t')} = {i E_{\beta}/\hbar \over 
\left\{1-\exp[i E_{\beta} \Delta /\hbar] \right\} \langle \phi_{\beta}
|{\partial \over \partial t} | \phi_{\alpha} \rangle }
\end{equation}
Using the fact that 
$\langle \phi_{\beta} | \partial/\partial t| \phi_{\alpha} \rangle = 
\langle \phi_{\beta} | \nabla_{\uR} | \phi_{\alpha} \rangle \cdot d\uR/dt = 
\uD_{\beta \alpha} \cdot \uv$, substituting into Eq.(\ref{eq:fpapprox2}), and
taking the limit as $\Delta \rightarrow 0$ we find that the Pechukas 
effective force at the beginning of the timestep can be approximated as
\begin{equation}
\uF_{\beta \alpha}^P(t') \sim -\left\{ {(E_{\beta} - E_{\alpha}) \over 
(\hat{\uD}_{\beta \alpha}\cdot \uv) \Delta} \right\} \hat{\uD}_{\beta \alpha}
\label{eq:cokerforce}
\end{equation}
where $\hat{\uD}_{\beta \alpha} = \uD_{\beta \alpha}/|\uD_{\beta \alpha}|$.
Every quantity appearing in this equation is evaluated at $t'$ so this
approximate Pechukas effective force is local in time due to our resolution of
the mixed state wavefunction into eigenstates at the beginning and end of the
step. The work done by this localized Pechukas force during $\Delta$ is just
\begin{equation} 
\int_{t'}^{t'+\Delta} \uF_{\beta \alpha}^P \cdot \uv dt \sim 
\uF_{\beta \alpha}^P \cdot \uv \Delta = -(E_{\beta}-E_{\alpha}) 
\end{equation}
the energy gap between the states so the dynamics is conservative. Further
\begin{equation} 
\int_{t'}^{t'+\Delta} \uF_{\beta \alpha}^P \cdot \uv dt = 
M \int_{t'}^{t'+\Delta} 
{d\uv \over dt} \cdot \uv dt = M \int_{\uv(t')}^{\uv(t'+\Delta)} \uv \cdot d\uv
\end{equation}
so from above
\begin{equation}
\frac{1}{2} M \uv^2(t'+\Delta) = \frac{1}{2} M \uv^2(t') 
- (E_{\beta}(t') - E_{\alpha}(t'))
\label{eq:econserve}
\end{equation}
{\em i.e.} the localized Pechukas effective force acts to redistribute the
electronic energy gap between the adiabatic states into the nuclear kinetic
energy.  The effectively impulsive dynamically localized Pechukas force in
Eq.(\ref{eq:cokerforce}) points in the $\hat{\uD}_{\beta \alpha}$ direction so
the only component of the velocity it can influence is parallel to this
vector. This is precisely the velocity readjustment algorithm which Tully and
others have used in nonadiabatic dynamics algorithms for some years
\cite{Tully1}. Our implementation of the force in Eq.(\ref{eq:cokerforce}) for
multidimensional systems is described in detail in Appendix A. The derivation
of the velocity readjustment procedure presented here is considerably simpler
than the earlier semiclassical justification given by Herman
\cite{Herman3,Herman4}.

\section{Results}
\setcounter{equation}{0}
\label{sec:results}
\baselineskip=24pt
\parindent=0.25in

In this section we compare the results of calculations performed 
using the surface hopping trajectory methods outlined in section
\ref{subsec:tully} with the results of semiclassical calculations using 
stationary phase trajectories described in section \ref{subsec:mixed_q-c}, as
well as with fully quantum calculations. The systems we study are the one
dimensional, two-state scattering potentials used as test problems by Tully
\cite{Tully3} and presented in Fig. \ref{fig:tully_pots}. These example
problems include both a single avoided crossing, and multiple successive
avoided crossing situations where accurate treatment of quantum interference
effects is crucial.  In our mixed quantum-classical calculations the nuclear
motion over these surfaces is treated classically and the 2x2 electronic
problem is handled quantum mechanically. Numerically exact, fully quantum
treatment of both the electronic and nuclear motion for these model problems
is easily obtained with wave packet propagation methods
\cite{Kosloff3,Coker12}. 

For all these calculations the initial wave function was a gaussian as 
in Eq.(\ref{eq:psi_init}) with $\sigma=2.5176au$, $R_0=-8au$ and we use 
$M=2000au$ as our nuclear mass \cite{Tully3}. In our surface 
hopping trajectory calculations we average results over an ensemble 
of 5000 trajectories whose starting positions were chosen according 
to the initial gaussian density, and all ensemble members were given 
the same initial momentum $M\dot{R} = \hbar k$. Our use of an 
ensemble enables us to integrate over the initial distribution of 
positions as well as over the different possible points at which 
stochastic hops between the electronic surfaces can take place. With 
our nonlocal semiclassical calculations, on the other hand, after 
specifying the electronic transition of interest we need only integrate 
over the initial position distribution as the dynamics  is completely 
deterministic. For these calculations, all trajectories which start in the 
asymptotic region with the same initial momentum will follow the 
same set of points in phase space. Each nonlocal semiclassical trajectory 
originates from a 
different initial position and thus its contribution to the average 
trajectory at time $t$ will simply be displaced in time from the 
others. In the nonlocal dynamics results reported below we thus use 
an ensemble of 300 points sampled from the initial gaussian 
distribution. The numerically exact full quantum solution was obtained using 
standard FFT propagation techniques \cite{Kosloff3,Coker12} with  a 4096 point 
spatial grid on the range -20. to $20au$. 

We compare properties such as branching ratios into the different coupled
adiabatic electronic states. Various average dynamical quantities are compared
from results computed using expressions like
Eqs.(\ref{eq:gen_av})-(\ref{eq:R_av}) with the Pechukas stationary phase
trajectories, averaged surface hopping trajectories, and the Ehrenfest
trajectories computed from fully quantum wave packet propagation results.  We
also compare the dynamical evolution of the position distributions computed
with the different methods ({\em cf.} Eq.(\ref{eq:prob_fin})).

The average phase space trajectories computed using these various 
methods for the coupled potentials in example 1 (see Fig. 
\ref{fig:tully_pots}a) with low initial momentum, {\em i.e.} trajectory energy 
just above the asymptotic excited state value, are presented in 
Fig.\ref{fig:ex1_k10}a. For this potential we find that the average 
trajectories obtained from the different mixed quantum-classical 
techniques agree quite well with the Ehrenfest trajectories calculated 
from full quantum wave packet propagation results. In all these 
calculations the quantum system was initially in state 1 and we 
compute the average position and momentum from the 
instantaneous distributions evolved over the different surfaces. 
There is no amplitude in the excited state until the coupling region is 
encountered so the averages over the excited state distributions are 
initially undefined. 

Fig.\ref{fig:ex1_k10}b displays the initial and final position distributions
computed from the squares of the components of the instantaneous nuclear wave
functions (full quantum), histograms of the surface hopping trajectory swarm
member positions, or a set of points sampled from the initial wave function
density, propagated according to the Pechukas equations of motion, and
weighted by their terminal transition probabilities according to
Eq.(\ref{eq:prob_fin}). We see that at this energy just above threshold for
the example 1 potentials, the mixed quantum-classical methods give dynamically
evolving distributions which are too strongly peaked compared with the full
quantum results. As discussed in the previous paragraph, the mean terminal
positions on the different surfaces are in excellent agreement with full
quantum results but the dispersion of the terminal distribution is too small
at this low momentum.  In Fig.\ref{fig:ex1_k10}c we present the time evolution
of the state occupation probabilities computed using the various methods
during this near threshold scattering event. We see that even though the mixed
quantum-classical distributions are too narrow, their areas give a quite
accurate representation of the dynamical branching between the coupled
electronic states.

When the initial momentum is such that the trajectory energy is lower than the
energy of the first excited state in the asymptotic region a number of effects
can cause deviations between results of the mixed quantum-classical methods
and the full quantum calculations: Transient trapping in the excited state
well can give rise to unphysical back reflected density on the ground state
\cite{Coker12,Tully3}. The full quantum wave function tunnels deep 
into the wall of the excited state well. Thus, as the quantum density rains
down onto the ground state through nonadiabatic interactions, it is still
traveling in the forward direction. With the mixed quantum-classical
calculations, on the other hand, the trajectories reach their turning points
in the excited state well and reverse their momenta.  Thus as swarm members
are coupled down onto the ground state they appear as back reflected density.
Related problems arise when the energy is lower than the ground state barrier
height as the mixed quantum-classical calculations cannot represent nuclear
tunneling. Similar effects give rise to the narrow mixed quantum-classical
distributions noted in Fig.\ref{fig:ex1_k10}b. At low momenta the classical
trajectories see sharp potential features where as the quantum nuclear
dispersion tends to average out these effects. Thus the mixed
quantum-classical trajectories tend to bunch up compared to the full quantum
results.

Results for the example 1 potentials with higher initial momentum $\hbar k =
M\dot{R}=25au$ are presented in Fig.\ref{fig:ex1_k25}. The near threshold
problems discussed above are no longer issues at these higher momenta and
excellent agreement between full quantum and mixed quantum-classical results
is observed for this example problem with a single region of nonadiabatic
interaction.

In Fig. \ref{fig:ex1_P} we present the probabilities of branching into the
different electronic states as functions of initial momentum for the example 1
potentials computed using the different mixed quantum-classical methods as
well as from full quantum wave packet propagation calculations. The agreement
between full quantum and mixed quantum-classical results is remarkably good
across the wide range of momenta considered for this example problem.

In this figure we also present the occupation probability normalization
$P_1+P_2$ as a function of time. The full quantum wave packet propagation
results conserve the norm very accurately indicating that our time step is
sufficiently small to give accurate integration of the equations of motion.
Since the total number of ensemble members in our surface hopping trajectory
calculations is a constant, and the branching of these members onto the
different surfaces is obtained from the time dependent quantum subsystem wave
function for each individual trajectory, these results conserve the norm by
construction. The nonlocal results presented here highlight the norm
conservation problems of Pechukas' stationary phase mixed quantum-classical
dynamics. As discussed in section \ref{subsec:mixed_q-c} the terminal
occupation probabilities of the different states are determined by solving the
quantal subsystem dynamics along different stationary phase paths which depend
on the terminal state. Thus the nonlocal semiclassical estimates of the
occupation probabilities are $P^{NL}_1[\bar{\uR}_{11}(t)]$ and
$P^{NL}_2[\bar{\uR}_{21}(t)]$ for our two state example problems considered
here. In Fig. \ref{fig:ex1_P} we see that for the example 1 potential surfaces
the sum $P_1^{NL} + P_2^{NL}$ remains less than unity across the entire
momentum range considered. For this problem with a single region of
nonadiabatic interaction we see that the deviation from normalization is never
more than a few percent. As noted in section \ref{subsec:mixed_q-c} these
deviations arise from the neglect of nonstationary paths in Pechukas'
formulation.

In Figs. \ref{fig:ex2_k16} - \ref{fig:ex2_k30} we present similar sets of
results to those discussed above for the example 2 potentials with various
initial momenta corresponding to $\hbar \uk=$ 16, 25, and 30$au$. The
qualitative shapes of the average mixed quantum-classical trajectories
obtained with either the local or nonlocal algorithms are all in reasonable
agreement with the full quantum Ehrenfest trajectories over these coupled
surfaces for this range of momenta.

The occupation probabilities presented in Figs. \ref{fig:ex2_k16}c-
\ref{fig:ex2_k30}c reveal the most significant deviations between the various 
mixed quantum-classical propagation methods and fully quantum results. For
$\hbar \uk = 16au$ we see that Tully's phase coherent surface hopping method
predicts too little branching onto state 2, at $\hbar \uk = 25au$ it predicts
to high a terminal population in state 2, and for $\hbar \uk = 30au$ this
surface hopping method fortuitously gets the branching just right. These
deviations are reflected in the sizes and shapes of the terminal position
distributions in Figs.\ref{fig:ex2_k16}b-\ref{fig:ex2_k30}b.

The occupation probabilities obtained from the nonlocal mixed
quantum-classical Pechukas dynamics algorithm show considerable norm
conservation problems for the example 2 potentials in that these dynamical
curves in Figs.\ref{fig:ex2_k16}c-\ref{fig:ex2_k30}c donot reflect about
$P=0.5$. Despite these problems, which we shall explore in more detail
shortly, we see that $P^{NL}_2[\bar{\uR}_{21}(t)]$ agrees remarkably well with
full quantum results across this entire range of initial momenta. On the other
hand, $P^{NL}_1[\bar{\uR}_{11}(t)]$ gives a rather poor representation of the
dynamical evolution of the occupation probability of state 1 over this
momentum range. In Fig. \ref{fig:ex2_k16_wave} we show in detail how the
nuclear densities on the different electronic states evolve through the
coupling region using our various propagation methods for $\hbar \uk = 16au$.
We see that the densities obtained from the nonlocal Pechukas theory results
for the $\bar{\uR}_{21}$ trajectories give a remarkably accurate description
of the evolution of the excited state nuclear wave function density
$|\chi_2|^2$. Even transient ripples in the excited state density are
reproduced with remarkable accuracy by the Pechukas stationary phase results.
Similar quality is found for these excited state results across the entire
range of momenta studied. The quality of the results for the ground state
density tends to follow the deviations in branching with different momentum.

Finally in Fig. \ref{fig:ex2_P} we present the terminal branching
probabilities predicted by the different methods as functions of initial
momentum for the example 2 potential surfaces. The most striking feature of
the results presented here is magnitude of the deviations from norm
conservation of the nonlocal Pechukas theory results.  We see the sum
$P_1^{NL} + P_2^{NL}$ deviates both above and below unity by more than 20\%
for some momenta. As discussed in section \ref{subsec:mixed_q-c} the
deviations below and above one result from the neglect of constructively and
destructively interfering nonstationary phase paths respectively, in the
Pechukas theory. Again we see that the branching probability $P_2^{NL}$
obtained from the $\bar{\uR}_{21}$ path is in excellent agreement with the
full quantum results, thus the norm conservation problems seem to result
entirely from the $P_1^{NL}$ values which deviate significantly from the full
quantum wave packet values. This suggests that for this problem, the
propagator $K_{21}$ is strongly dominated by the stationary phase contribution
from the single path $\bar{\uR}_{21}$. The ground state propagator $K_{11}$,
however, is only approximately represented by the contribution from the
$\bar{\uR}_{11}$ path, and there must also be significant contributions
arising from interferences between nonstationary paths which are neglected in
our calculations.

From Fig. \ref{fig:ex2_P} we see that the branching probabilities obtained
with Tully's algorithm for this problem ofcourse give good norm conservation
and they generally show absolute deviations from the full quantum results
which are at worst no more than about 10\%.

\section{Conclusion}
\setcounter{equation}{0}
\label{sec:conclusion}

In summary we have presented a detailed study of the shortcomings and
relationships between various methods for mixed quantum-classical propagation.
Using time dependent perturbation theory we have shown that local surface
hopping dynamics methods can be derived from the more rigorous nonlocal
equations of motion obtained by Pechukas from a stationary phase approximation
to the reduced propagators. By extending Pechukas' stationary phase approach
to compute state occupation probabilities we have developed an understanding
of the precise relationship between classical path initial conditions and the
properties of the initial fully quantum wave packet being modeled.

Further we have shown that the semiclassical reduced propagators obtained as
sums over the Pechukas stationary phase paths in general give a full system
dynamics which does not conserve the norm. In fact we have seen that,
depending on the problem, some elements of the reduced propagator can be
accurately approximated by including only the stationary phase paths, whilst
other elements have significant contributions from nonstationary paths for
which the phase is probably varying quite slowly. Under such circumstances the
detailed phase cancellation of many nonstationary paths needs to be considered
and path sampling methods for treating such problems need to be more fully
developed in the context of nonadiabatic dynamics. It may prove interesting to
explore the properties of paths for different transitions whose phase has been
optimized subject to the constraint that the norm is conserved. This builds
norm conservation into the phase optimization procedure in such a way as to
make paths for different transitions dependent upon one another.

The accuracy of phase coherent local surface hopping dynamics methods is
encouraging and it may prove fruitful to use the swarms of paths produced in
these calculations as starting points to correct for the temporal localization
inherent in these methods.

\section{Acknowledgements}
We gratefully acknowledge financial support for this work from the National
Science Foundation (Grant No. CHE-9058348), and the Petroleum Research Fund
administered by the American Chemical Society (Grant No. 27995-AC6), and a
generous allocation of supercomputer time from the National Center for
Supercomputing Applications.

\newpage

\section{Appendix A}
\label{sec:summary}

For general applications we follow Webster {\em et al.} \cite{Webster1} and
assume that the quantum subsystem hamiltonian varies linearly with time over a
nuclear timestep $\Delta$. In our application this assumption serves simply as
a computational convenience to reduce the frequency with which we compute the
dynamically changing basis set. If the interaction between the quantal and
classical subsystems is described by a potential ${\cal V}(\ur,\uR(t))$ then
our first guess at the time dependence of the quantal hamiltonian over the
interval $t \rightarrow t+\Delta$ is to assume that the linear variation of
the potential over the current step is the same as over the previous step.
Thus we break the classical timestep $\Delta$ up into $N_s$ small steps
$\delta = \Delta/N_s$ to evolve the forward propagating quantal expansion
coefficients. Our initial prediction for the time dependent quantal
hamiltonian matrix at the $l^{\rm th}$ small timestep in the interval
$t\rightarrow t+\Delta$ is thus
\begin{equation}
{\cal H}_{ij}^p(t+l\delta) = E_i(t) \delta_{ij} + l\delta \langle \phi_i(t) |
\Delta{\cal V}^p(t) | \phi_j(t) \rangle /\Delta \label{eq:linhamp}
\end{equation}
where $\Delta{\cal V}^p = {\cal V}(\ur,\uR(t)) - {\cal V}(\ur,\uR(t-\Delta))$
and $\phi_i(t)$ and $E_i(t)$ are the instantaneous adiabatic eigenstates and
energies of the quantal subsystem for the current nuclear configuration
$\uR(t)$. Diagonalizing the linear approximate predicted hamiltonian in
Eq.(\ref{eq:linhamp}) gives an efficient means of obtaining approximate
eigenstates at intermediate times $t+l\delta$ which can then be used as a
basis set to accurately integrate the expansion coefficient equations of
motion Eq.(\ref{eq:amotion}). Our algorithm thus proceeds as follows:

\begin{description}
\item[\rm (i)]  We use the linear extrapolated hamiltonian in
Eq.(\ref{eq:linhamp}) together with a split operator electronic propagator
\cite{Webster1} (see appendix B) to integrate the forward propagating expansion
coefficients coherently from their current values at time $t$ and predict
their values at $t+\Delta$. At each small timestep $t+l\delta$ during the
predictor stage we compute the Tully hopping probabilities $g_{\alpha
\rightarrow \beta}$ from Eq.(\ref{eq:gitok}) and attempt hops. Hops are only
allowed which give real velocities thus we only allow state switching for
trajectories whose total energies (instantaneous nuclear kinetic energy plus
current occupied adiabatic state energy) are greater than the adiabatic energy
of the new state to be occupied as a result of the transition.

\item[\rm (ii)] If no hop is successful during the $N_s$ small steps between
$t$ and $t+\Delta$ during the predictor cycle then the localized Pechukas
force on the nuclei at time $t$ is obtained by substituting $\beta = \alpha$
into Eq.(\ref{eq:fpapprox1}) in which case the localized Pechukas effective
force is simply the usual Hellman-Feynman force
\begin{equation}
\uF_{\alpha \alpha}^P = - \nabla_{\uR} E_{\alpha}
\end{equation} 
and we integrate the nuclear motion for a step with the velocity Verlet
integrator. 

\item[\rm (iii)] If a hop has occured during the predictor step then the 
localized Pechukas effective force given in Eq.(\ref{eq:cokerforce}) must be
used to evolve the nuclear equations of motion through $\Delta$. We do this by
first predicting the velocities at time $t$ according to the following result
which uses the acceleration associated with the localized Pechukas force in
Eq.(\ref{eq:cokerforce})
\begin{equation}
\uv^p(t) = \uv(t-\Delta) -\left\{ {(E_{\beta} - E_{\alpha}) \over 
\hat{\uD}_{\beta \alpha}\cdot \uv(t-\Delta)} \right\} 
{\hat{\uD}_{\beta \alpha} \over M}
\end{equation}
This predicted velocity is changed most strongly in the direction of the
nonadiabatic coupling vector $\hat{\uD}_{\beta \alpha}$. The factor
\begin{equation}
s = \left[{ 2(E^{traj} - E_{\beta}(t)) \over M \uv^p(t) \cdot \uv^p(t) }
\right]^{\frac{1}{2}}
\end{equation}
is then computed and all the particle velocities are rescaled so as to give
exact energy conservation {\em i.e.} $\uv(t) = s\uv^p(t)$. Here $E^{traj}$ is
the total energy of the given trajectory. The new position is then computed
from
\begin{equation} 
\uR(t+\Delta) = \uR(t) +\uv(t) \Delta + {\Delta^2 \over 2M}
\uF_{\beta \beta}^P(t)
\label{eq:verletr} 
\end{equation} 

\item[\rm (v)] Finally we compute the instantaneous adiabatic eigenstates and
energies at the new configuration $\uR(t+\Delta)$ and use the following linear
approximate time dependent hamiltonian
\begin{equation}
{\cal H}_{ij}^c(t+l\delta) = E_i(t) \delta_{ij} + l\delta \langle \phi_i(t) |
\Delta{\cal V}^c(t) | \phi_j(t) \rangle /\Delta \label{eq:linhamc}
\end{equation}
where $\Delta{\cal V}^c = {\cal V}(\ur,\uR(t+\Delta)) - {\cal V}
(\ur,\uR(t))$ to reintegrate and correct the forward propagating expansion
coefficients giving accurate coherent integration of the forward propagating
mixed electronic state for all times. 

\end{description}

The algorithm outlined in this Appendix is set up in an efficient way so that
it can be used to study large scale systems for which obtaining the
instantaneous basis sets to describe the interacting electronic states is the
most computationally intensive part of the calculation. As we are propagating
an ensemble of independent surface hopping trajectories, considerable
computational gains can be obtained by taking advantage of the obvious
parallelism in these calculations. Applications of this algorithm to study
nonadiabaticity in excited state condensed phase chemical reactions will be
reported in future publications.

\section{Appendix B}
\label{sec:pechukas_details}

Pechukas' self-consistent stationary phase trajectories are relatively
straight forward to compute for these simple 1D, two state problems. First we
specify the initial position and momentum of the classical trajectory, the
initial and final quantum states of interest, and finally the time, $T$, for
which the trajectory is to be followed.  No final point information about the
classical coordinates is required. Next we break the time interval up into $N$
classical steps of length $\Delta$ and choose positions for the classical
coordinates at each time point providing an initial guess at the nuclear
trajectory. For the Tully examples considered here, we have chosen a simple
linear trajectory with constant velocity as our initial guess. This guessed
trajectory defines a time dependent quantal Hamiltonian which we use to
compute the quantal subsystem propagator which evolves the mixed state quantum
wavefunction forward and backward in time away from the specified initial and
final boundary states.

We find it convenient to work with instantaneous adiabatic basis states whose
coefficient vector $\ua(t)$ is readily propagated using
\begin{equation}
\ua(t+\delta) = \uT(t+\delta,t) \ua(t)
\end{equation}
where the propagator matrix elements are easily constructed from the
instantaneous eigenvalues $E_n(t)$ and eigenfunctions $\phi_n(t)$ using the
following short-time approximation
\begin{equation}
\uT_{no}(t+\delta,t) = \exp\left[{-i \delta \over 2\hbar}\left(E_n(t+\delta) 
+E_o(t) \right) \right] \langle \phi_n(t+\delta)|\phi_o(t) \rangle
\end{equation}
Linear interpolation of the Hamiltonian matrix elements makes for efficient
implementation of this algorithm as discussed in Appendix A.

With these forward and backward propagated mixed state coefficients evaluated
at all the classical time points we compute the Pechukas forces on the
trajectory at these times from Eq.(\ref{eq:fp_adia}) and use a simple velocity
Verlet integrator to solve the Pechukas equation of motion,
Eq.(\ref{eq:pechmot}), giving a new trajectory. This process is iterated till
the trajectory converges as described in section \ref{subsec:semic}. For some
momenta with the example 2 potentials we have found it difficult to get the
$\uR_{21}$ trajectory to converge with the simple iteration scheme discussed
here (see Fig. \ref{fig:ex2_P}). We are currently exploring the reasons for
these problems.

\newpage
     
\section{Figure Captions}
\begin{description}

\item[Fig.~\ref{fig:int_paths}] Generalized path variation scheme including 
endpoint variations. The structure of the paths which contribute to the
integrand in Eq.(\ref{eq:prob_sc}) involve two different starting points
$\uR'$ and $\uR'''$ (x's) connected by two different stationary paths (solid
lines) to a common terminal point $\uR''$. The open circles are varied path
endpoints connected by new stationary phase paths (dashed lines). Forward and
backward path variation functions are indicated by the arrows.

\item[Fig.~\ref{fig:tully_pots}] One dimensional, two-state example potential
energy curves taken from ref. \cite{Tully3}

\item[Fig.~\ref{fig:ex1_k10}] Results from various propagation schemes for 
example 1 potentials. Initial gaussian nuclear wave packet starts on the
ground electronic state with momentum $\hbar \uk=10au$. In all figures full
quantum wave packet results are the solid curves, dashed curves are Pechukas
theory results, and dotted curves are surface hopping calculations. (a) Mean
phase space trajectories ($\langle \uk \rangle$ v's $\langle \ux \rangle$).
Curves labeled $\chi_1$ and $\chi_2$ are averaged over the instantaneous
ground and excited state nuclear distributions respectively. The adiabatic
potential energies are also presented as the bottom two short dashed lines.
The short dashed line showing the constant total energy for this trajectory is
indicated by the arrow. (b) Initial ($|f|^2$) and final ($|\chi_1|^2$,
$|\chi_2|^2$) wave function densities. (c) Time evolution of state occupation
probabilities $P_1$ and $P_2$.

\item[Fig.~\ref{fig:ex1_k25}] Same as Fig. \ref{fig:ex1_k10} except initial 
momentum is $\hbar \uk=25au$.

\item[Fig.~\ref{fig:ex1_P}] Occupation probabilities as functions of initial 
momentum for example 1 potentials using various propagation schemes.  Solid
curves give full quantum results, Pechukas theory values are the dashed
curves, and the dotted curves give surface hopping results. We also display
the sum $P_1 + P_2$ to highlight norm conservation problems.

\item[Fig.~\ref{fig:ex2_k16}] Results for various propagation schemes for 
example 2 potentials. Initial gaussian nuclear wave packet starts on the
ground electronic state with momentum $\hbar \uk=16au$. In all figures full
quantum wave packet results are the solid curves, dashed curves are Pechukas
theory results, and dotted curves are surface hopping calculations. (a) Mean
phase space trajectories ($\langle \uk \rangle$ v's $\langle \ux \rangle$).
Curves labeled $\chi_1$ and $\chi_2$ are averaged over the instantaneous
ground and excited state nuclear distributions respectively. The adiabatic
potential energies are also presented as the bottom two short dashed lines.
The short dashed line showing the constant total energy for this trajectory is
indicated by the arrow. (b) Initial ($|f|^2$) and final ($|\chi_1|^2$,
$|\chi_2|^2$) wave function densities. (c) Time evolution of state occupation
probabilities.

\item[Fig.~\ref{fig:ex2_k25}] Same as Fig. \ref{fig:ex2_k16} except initial 
momentum is $\hbar \uk=25au$.

\item[Fig.~\ref{fig:ex2_k30}] Same as Fig. \ref{fig:ex2_k16} except initial 
momentum is $\hbar \uk=30au$.

\item[Fig.~\ref{fig:ex2_k16_wave}] Details of wave packet evolution through 
the coupling region on example 2 potential surfaces for initial momentum of
$\hbar \uk = 16 au$. Full quantum results are presented as solid curves,
dashed curves are Pechukas theory results, and dotted curves are surface
hopping calculations.  (a) Instantaneous ground state nuclear densities at
various times, and (b) gives corresponding excited state densities.

\item[Fig.~\ref{fig:ex2_P}] Occupation probabilities as functions of initial 
momentum for example 2 potentials using various propagation schemes.  Solid
curves give full quantum results, Pechukas theory values are the dashed
curves, and the dotted curves give surface hopping results. We also display
the sum $P_1 + P_2$ to highlight norm conservation problems. At intermediate
momenta we found it difficult to converge the Pechukas stationary phase
iteration process, the gaps in the curves reflect these problems.

\end{description}
     
\newpage
     
\bibliography{nonad_refs}
\bibliographystyle{unsrt}
    
\newpage
 
\begin{figure}
\caption{integration paths}
\label{fig:int_paths}
\end{figure}

\begin{figure}
\caption{tully example pots}
\label{fig:tully_pots}
\end{figure}

\begin{figure}
\caption{example 1 k=10}
\label{fig:ex1_k10}
\end{figure}

\begin{figure}
\caption{example 1 k=25}
\label{fig:ex1_k25}
\end{figure}

\begin{figure}
\caption{example 1 occupation probabilities}
\label{fig:ex1_P}
\end{figure}

\begin{figure}
\caption{example 2 k=16}
\label{fig:ex2_k16}
\end{figure}

\begin{figure}
\caption{example 2 k=25}
\label{fig:ex2_k25}
\end{figure}

\begin{figure}
\caption{example 2 k=30}
\label{fig:ex2_k30}
\end{figure}

\begin{figure}
\caption{example 2 wave function propagation}
\label{fig:ex2_k16_wave}
\end{figure}

\begin{figure}
\caption{example 2 occupation probabilities}
\label{fig:ex2_P}
\end{figure}

\end{document}